# Exact Evaluation of The Resistance in an Infinite Face- Centered Cubic Network

Jihad H. Asad

Dep. of Physics- Tabuk University- P. O Box 741- Tabuk 71491- Saudi Arabia

**Abstract**

The equivalent resistance between the origin and the lattice site (2n,0,0), in an infinite Face Centered Cubic network consisting from identical resistors each of resistance *R*, has been expressed in terms of the complete elliptic integral of the first kind, and $\pi$. The asymptotic behavior is investigated, and some numerical values for the equivalent resistance are presented.

# I-Introduction

The Lattice Green Function (LGF) plays a key role in the theory of solid state physics. In literature, many efficient approaches have been reported for these functions. As seen from literature most of the studies on lattice functions are based on elliptic integral and recurrence relation approaches [1-11]. As it is noticed many quantities of interest in solid state physics can be expressed in terms of LGF, for example, statistical model of ferromagnetism such as Ising model [12], Heisenberg model [13, and spherical model [14]; lattice dynamics [15], random walk theory [11,16], and band structure [17].

The LGF for the FCC lattice has been studied and investigated well. Watson [18] showed that:

$$F(3;0,0,0) = f_o = \frac{\sqrt{3}}{\pi^2}[K(k)]^2 = \frac{3\Gamma^6(\frac{1}{3})}{2^{\frac{14}{3}}\pi^4} = 0.4482203944. \qquad (1)$$

where $k = Sin\frac{\pi}{12} = \frac{\sqrt{3}-1}{2\sqrt{2}}$, and it is equal to the singular modulus $k_3$.

As explained in Inoue [5] the LGF for the FCC lattice at any site with nearest neighbor interaction can be, in principle, expressed in terms of the complete elliptic integrals of the first and second kind.

Morita [7] showed that the LGF for FCC lattice at any arbitrary lattice site can be calculated from the three known values at the lattice sites $(0,0,0)$, $(2,0,0)$, and $(2,2,0)$ using the so-called recurrence formula presented in Inoue paper [5].

Morita and Horighuchi [6] explain how to evaluate he LGF for many lattices, where the formula they introduced involve the complete elliptic integral of the first kind. On the other hand, the LGF for cubic lattices (i.e., SC, BCC, and FCC lattices) can be expressed rationally in terms of $\pi$ and the known value of the LGF at the origin [19].

Finally, Joyce, and Delves [20] studied the mathematical properties of the LGF for the FCC lattice, where they present a relation for calculating both $F(2n,0,0;w)$, and $F(2n,0,0;3)$, in addition to studying the asymptotic behavior as $n \to \infty$.

The computation of the resistance between two nodes in a resistor network is an interesting classic problem in the electric circuit theory, where it attracts the attention of many authors over many years. Besides being a central problem in electric circuit theory, the computation of resistances is also relevant to a wide range of problems ranging from random walk [21,22], theory of harmonic functions [23], to first-passage processes [24], to Lattice Green's Functions [25]. The connection with these problems originates from the fact that electrical potentials on a grid are governed by the same difference equations as those occurring in the other problems. For this reason, the resistance problem is often studied from the point of view of solving the difference equations, which is most conveniently carried out for infinite networks. Kirchhoff [26] formulated the study of electric networks more than 150 years ago. The electric circuit theory is discussed in detail in a classic text by Van der Pol and Bremmer[27] where they derived the resistance between nearby points on the square lattice.

Later on, many efforts have been focused on analyzing infinite networks of identical resistors, where many approaches have been used. The superposition of current distribution has been used to calculate the effective resistance between adjacent sites on infinite networks [28-30]. The mathematical problem of this method involves the solution of an infinite set of linear, inhomogeneous difference equations which are solved by the method of separation of variables. Numerical results for the resistances between arbitrary sites are presented.

A mapping between random walk problems and resistor networks problems have been used by Monwhea Jeng [31], where his method was used to calculate the effective resistance between any two sites in an infinite two-dimensional square lattice of unit resistors and the superposition principle was used to find the effective resistances on toroidal- and cylindrical-square lattices.

A third important method based on the LGF of the lattices have been introduced by Cserti [32,33], this method can be used to study both perfect and perturbed infinite networks, and many papers have been published using this method [34-38].

Finally, Wu [39] introduced a method which enables one to study finite networks, where the equivalent resistance can be obtained in terms of the eigenvalues and eigenfunctions of the Laplacian matrix associated with the network.

This work is oriented as follows: In Sec. II, we briefly discussed the basic formulas for the LGF of the FCC network. In Sec. II, an application to the LGF of the FCC network has been applied to calculate the equivalent resistance between the origin and the lattice site (2n,0,0) in the infinite FCC network. We close this work (i.e., Sec. IV) with a discussion to the results obtained in this study.

## II-General Relations for The LGF of The FCC Lattice

The Lattice Green Function for the FCC lattice which appears in many statistical problems [1-5], is defined as [1]:

$$F(n,m,l;w) = \frac{1}{\pi^3} \int_0^\pi \int_0^\pi \int_0^\pi \frac{Cos n\theta_1 Cos m\theta_2 Cos l\theta_3}{w - (Cos\theta_1 Cos\theta_2 + Cos\theta_2 Cos\theta_3 + Cos\theta_1 Cos\theta_3)} d\theta_1 d\theta_2 d\theta_3 .$$

where $n+m+l = $ even integer, and $w = w_1 + iw_2$ is a complex variable.

The triple integral given in Eq.(1) above was evaluated by Watson [18], for the special case where $n = m = l = 0$, and $w = 3$, where he got the following result:

$$F(0,0,0;3) = \frac{\sqrt{3}}{\pi^2}\left[K\left(\frac{\sqrt{3}-1}{2\sqrt{2}}\right)\right]^2 = \frac{3}{2^{\frac{14}{3}}\pi^4}\left[\Gamma\left(\frac{1}{3}\right)\right]^6 = 0.4482203944 . \quad (2)$$

where $K(k)$ is the complete elliptic integral of the first kind with a modulus $k$.

The above special value is called LGF of the FCC at the origin, and from now on we will refer to it throughout this paper as $f_o$ (i.e., $f_o = F(0,0,0;3)$). A general expression has been given by Iwata [8] where he expressed $f_o$ as:

$$F(0,0,0;w) = \frac{4}{\pi^2(w+1)}K(k_+)K(k_-) . \quad (3)$$

where $k_\pm \equiv k_\pm(w) = \frac{1}{2} \pm \frac{2}{w}(1+\frac{1}{w})^{\frac{-1}{2}} - \frac{1}{2}(1-\frac{1}{w})(1+\frac{1}{w})^{\frac{-3}{2}}(1-\frac{3}{w})^{\frac{1}{2}} . \quad (*)$

It has been also showed by Joyce [ ] that:

$$\frac{K(k_+)}{K(k_-)} = (1+\frac{1}{w})^{\frac{-1}{2}}\left[2 - \left(1-\frac{3}{w}\right)^{\frac{1}{2}}\right] . \quad (4)$$

Now, on applying Eq. (4) to Eq. (3) one gets the following simplified expression:

$$F(0,0,0;w) = \frac{4}{\pi^2 w}(1+\frac{1}{w})^{\frac{-3}{2}}\left[2-\left(1-\frac{3}{w}\right)^{\frac{1}{2}}\right][K(k_-)]^2. \tag{5}$$

It has been showed by Inoue [5] that one can calculates $F(n,m,l;w)$ at any arbitrary lattice point using the so-called recurrence relation provided that the following $\{F(2n,0,0;w); n=0,1,2,...\}$, and $F(2,2,0;w)$ values are known. Also, he showed that $F(2n,0,0;w)$ can be expressed in terms of the hypergeometric function (i.e., $F_4$), of Appell's type.

From the formula presented in Inoue's paper [5] (i.e., Eq. 3.9) he deduced that $F(2n,0,0;w)$ for $n=0,1,2,...$ can be expressed and evaluated in terms of $K(k_\pm)$, and $E(k_\pm)$ where $K(k)$, and $E(k)$ are the compelte elliptic integral of the first and second kind of modulus $k$ respectively.

The triple sum obtained by Inoue [5] (i.e., Eq. 3.9) was later simplified by Mano [40], where it was found that:

$$F(2n,0,0;w) = \frac{1}{(w+1)^{2n+1}} \frac{\left(\frac{1}{2}\right)_n^2}{(n!)^2} {}_2F_1\left(n+\frac{1}{2}, n+\frac{1}{2}, n+1; k_+^2\right)$$

$$x \, {}_2F_1\left(n+\frac{1}{2}, n+\frac{1}{2}, n+1; k_-^2\right). \tag{6}$$

Where $k_\pm^2(w)$ is defined in Eq. (4) above.

More generally, it was showed [ ] that $F(n,m,l;w)$ can be determined from the knowledge of $F(0,0,0;w)$, $F(2,0,0;w)$, and $F(2,2,0;w)$.

In a recent work Joyce [7] showed that $F(2n,0,0;w)$ can be written as:

$$F(2n,0,0;w) = \frac{1}{(w+1)} \frac{\left(\frac{1}{2}\right)_n^2}{(n!)^2} \left[\frac{1}{2}\sqrt{w+1}-\sqrt{w-3}\right]^{4n}$$

$$x \, {}_2F_1\left(\frac{1}{2},\frac{1}{2}, n+1; k_+^2\right) {}_2F_1\left(\frac{1}{2},\frac{1}{2}, n+1; k_-^2\right). \tag{7}$$

Using the following Euler transformation:

$${}_2F_1\left(\frac{1}{2},\frac{1}{2}, n+1; k^2\right) = (1-k^2)^n \, {}_2F_1\left(n+\frac{1}{2}, n+\frac{1}{2}, n+1; k^2\right). \tag{8}$$

Then, Eq. (7) above reduces to Mano's formula (i.e., Eq. (6)) above.

Now, on substituting $w = 3$ into Eq. (7), and Eq. (*) one gets the following product form [20]:

$$F(2n,0,0;3) = \frac{1}{4} \frac{\left(\frac{1}{2}\right)_n^2}{(n!)^2} {}_2F_1\left(\frac{1}{2},\frac{1}{2};n+1;(k_3')^2\right) {}_2F_1\left(\frac{1}{2},\frac{1}{2};n+1;(k_3)^2\right). \tag{9}$$

where $k_3 = \frac{\sqrt{2}}{4}(\sqrt{3}-1)$. (10)

is the singular value of order three (refer to Ref. [3], page 139), and $k_3'$ is the complementary modulus.

The above can be rewritten as (i.e., see Joyce [20]):

$$F(2n,0,0;3) = (-1)^n \frac{\sqrt{3}}{3^n} \left\{ \left[\frac{\breve{U}_n^{(1)} K_3}{\pi}\right]^2 - \left[\frac{\breve{U}_n^{(2)}}{K_3}\right]^2 \right\}. \tag{11}$$

Where $\{\breve{U}_n^{(j)} : j = 1,2\}$ are rational numbers satisfying the following recurrence relation:

$$(2n+1)\breve{U}_{n+1}^{(j)} - 12n\breve{U}_n^{(j)} - 3(2n-1)\breve{U}_{n-1}^{(j)} = 0. \tag{12}$$

With $n = 1,2,...,$ and the following initial conditions $\breve{U}_0^{(1)} = 1, \breve{U}_1^{(1)} = 1, \breve{U}_0^{(2)} \equiv 0$, and $\breve{U}_1^{(2)} = 1$.

and with $K_3 = \frac{\pi}{2} {}_2F_1(\frac{1}{2},\frac{1}{2};1;k_3^2)$

Using Eq. (11) and Eq. (12) with the initial conditions one can gets the value of $f_o$ where

$$f_o = F(0,0,0;3) = \left(\sqrt{3}\left\{\left[\frac{1 K_3}{\pi}\right]^2 - \left[\frac{0}{K_3}\right]^2\right\}\right) = \frac{\sqrt{3}}{\pi^2} K_3^2. \tag{13}$$

### III-Application: Evaluation of the resistance R(2n,0,0) in An Infinite FCC Network

Now consider an infinite FCC network consisting of identical resistors each of resistance $R$. The aim of this section is to calculate the equivalent resistance between the origin and the lattice site $(2n,0,0)$ where $n = 0,1,2,3,...$
First of all, It has been showed that in general for a d-dimensional infinite network consisting of identical resistors each of resistance $R$, then the equivalent resistance between the origin and any other lattice site can be given as [32]:

$$R(l_1,l_2,...,l_d) = R\int_{-\pi}^{\pi}\frac{dx_1}{2\pi}...\int_{-\pi}^{\pi}\frac{dx_d}{2\pi}\frac{1-\exp(il_1x_1+il_2x_2+...+il_dx_d)}{\sum_{i=1}^{d}(1-Cosx_i)}. \quad (14)$$

On the other hand, the LGF for a d-dimensional hypercube can be written as [1]:

$$G(l_1,l_2,...,l_d) = \int_{-\pi}^{\pi}\frac{dx_1}{2\pi}...\int_{-\pi}^{\pi}\frac{dx_d}{2\pi}\frac{\exp(il_1x_1+il_2x_2+...+il_dx_d)}{2\sum_{i=1}^{d}(1-Cosx_i)}. \quad (15)$$

From the above two equations one the equivalent resistance can be expressed as [32]:

$$R(\vec{r}) = 2R[G(0) - G(\vec{r})]. \quad (16)$$

Where  $\vec{r} = l_1\vec{a}_1 + l_2\vec{a}_2 + ... + l_d\vec{a}_d$

$l_1, l_2,...,l_d$ are integers (positive, negative or zero),

and $\vec{a}_1, \vec{a}_2,...,\vec{a}_d$ are independent primitive translation vectors.

For our infinite FCC network the above equation can be rewritten as:

$$R(n,m,l) = R[f_o - F(n,m,l)]. \quad (17)$$

Where we have used $\vec{r} = n\vec{a} + m\vec{a} + l\vec{a}$

In this work we are interested in calculating the equivalent resistance between the origin and the lattice site $(2n,0,0)$ in the infinite FCC network. To do this we substitute Eq. (1), Eq. (11) into Eq. (17), we got:

$$R(2n,0,0) = \left(\frac{\sqrt{3}}{\pi^2}[K(k_3)]^2 - (-1)^n \frac{\sqrt{3}}{3^n}\left\{\left[\frac{\breve{U}_n^{(1)} K_3}{\pi}\right]^2 - \left[\frac{\breve{U}_n^{(2)}}{K_3}\right]^2\right\}\right). \tag{18}$$

Now, using Eq. (12) and Eq. (18) with the initial conditions of $\{\breve{U}_n^{(j)} : j = 1,2\}$, one can calculate the required equivalent resistance. Since the LGF is an even function (i.e., $F(2n,0,0;3) = F(-2n,0,0;3)$) then as a result the resistance also is symmetric due to the fact that the FCC network is pure and symmetric, and as a result $R(2n,0,0) = R(-2n,0,0)$. In the following we give some examples:

1- Taking $n = 0$, and using $\breve{U}_0^{(1)} = 1$, and $\breve{U}_0^{(2)} \equiv 0$. One, gets:

$$R(0,0,0) = \left(\frac{\sqrt{3}}{\pi^2} K_3^2 - \sqrt{3}\left\{\left[\frac{K_3}{\pi}\right]^2 - \left[\frac{0}{K_3}\right]^2\right\}\right) = 0.$$

2- Taking $n = 1$, using $\breve{U}_1^{(1)} = 1$, and $\breve{U}_1^{(2)} = 1$. One, gets:

$$R(2,0,0) = \left(\frac{\sqrt{3}}{\pi^2} K_3^2 - \left(-\frac{\sqrt{3}}{3}\left\{\left[\frac{K_3}{\pi}\right]^2 - \left[\frac{1}{K_3}\right]^2\right\}\right)\right) = \frac{4}{\sqrt{3}\pi^2} K_3^2 - \frac{1}{\sqrt{3}} \frac{1}{K_3^2}.$$

3- Taking $n = 2$, with $\breve{U}_2^{(1)} = 5$, and $\breve{U}_2^{(2)} = 4$. One, gets:

$$R(4,0,0) = \left(\frac{\sqrt{3}}{\pi^2} K_3^2 - \frac{1}{3\sqrt{3}}\left[\frac{25 K_3^2}{\pi^2} - \frac{16}{K_3^2}\right]\right) = \frac{-16}{3\sqrt{3}\pi^2} K_3^2 + \frac{16}{3\sqrt{3}} \frac{1}{K_3^2}.$$

4- Taking $n = 3$, with $\breve{U}_3^{(1)} = \frac{129}{5}$, and $\breve{U}_3^{(2)} = 21$. One, gets:

$$R(6,0,0) = \left(\frac{\sqrt{3}}{\pi^2} K_3^2 - \left[-\frac{\sqrt{3}}{27}\left\{\left[\frac{129 K_3}{5\pi}\right]^2 - \left[\frac{21}{K_3}\right]^2\right\}\right]\right) = \frac{1924\sqrt{3}}{75\pi^2} K_3^2 - \frac{49\sqrt{3}}{3} \frac{1}{K_3^2}.$$

5- Taking $n = 4$, with $\breve{U}_4^{(1)} = \frac{717}{5}$, and $\breve{U}_4^{(2)} = \frac{816}{7}$. One, gets:

$$R(8,0,0) = \left(\frac{\sqrt{3}}{\pi^2} K_3^2 - \frac{\sqrt{3}}{81}\left\{\left[\frac{717 K_3}{5\pi}\right]^2 - \left[\frac{816}{7 K_3}\right]^2\right\}\right) = \frac{-56896\sqrt{3}}{225\pi^2} K_3^2 + \frac{73984\sqrt{3}}{441} \frac{1}{K_3^2}.$$

6- Taking $n = 5$, with $\breve{U}_5^{(1)} = \frac{825}{1}$, and $\breve{U}_5^{(2)} = \frac{4695}{7}$. One, gets:

$$R(10,0,0) = \left(\frac{\sqrt{3}}{\pi^2}K_3^2 - \left[-\frac{\sqrt{3}}{243}\left\{\left[\frac{825K_3}{\pi}\right]^2 - \left[\frac{4695}{7K_3}\right]^2\right\}\right]\right) = \frac{75652\sqrt{3}}{27\pi^2}K_3^2 - \frac{2449225\sqrt{3}}{1323}\frac{1}{K_3^2}$$

7- Taking $n = 6$, with $\breve{U}_6^{(1)} = \frac{266859}{55}$, and $\breve{U}_6^{(2)} = \frac{27612}{7}$. One, gets:

$$R(12,0,0) = \left(\frac{\sqrt{3}}{\pi^2}K_3^2 - \frac{\sqrt{3}}{729}\left\{\left[\frac{266859K_3}{55\pi}\right]^2 - \left[\frac{27612}{7K_3}\right]^2\right\}\right) = \frac{949516345\sqrt{3}}{29403\pi^2}K_3^2 + \frac{9412624\sqrt{3}}{441}\frac{1}{K_3^2}$$

8- Taking $n = 7$, with $\breve{U}_7^{(1)} = \frac{1593171}{55}$, and $\breve{U}_7^{(2)} = \frac{2142999}{91}$. One, gets:

$$R(14,0,0) = \left(\frac{\sqrt{3}}{\pi^2}K_3^2 - \left[-\frac{\sqrt{3}}{2187}\left\{\left[\frac{1593171K_3}{55\pi}\right]^2 - \left[\frac{2142999}{91K_3}\right]^2\right\}\right]\right)$$

$$= \frac{1253432321\sqrt{3}}{3267\pi^2}K_3^2 + \frac{238111\sqrt{3}}{7371}\frac{1}{K_3^2}.$$

The above values can be calculated numerically by means of Mathematica. In Table 1 below we present some numerical calculated values for the resistance between the origin and the site *(2n,0,0)* are presented.

The asymptotic case (i.e., the separation between the origin and the site *(2n,0,0)* goes to a large value or infinity). In this case the resistance goes to a finite value. To explain this point, we note that from the theory of Fourier series (Riemann's Lemma) that $Lim_{n\to\infty}\int_a^b \Phi(x)Cosnxdx \to 0$ for any integrable function $\Phi(x)$. Thus, $F(n,m,l) \to 0$ (corresponding to the boundary condition of Green's function at infinity), and as a result Eq. (17) becomes

$$\frac{R_o(3,2n,0,0)}{R} \to f_0(3,0,0,0). \tag{16}$$

when $n \to \infty$.

**IV-Results and Discussion**

We have expressed the equivalent resistance between the origin and the lattice site *(2n,0,0)* in an infinite FCC network consisting of identical resistors each of resistance *R*. The equivalent resistance is obtained in terms of the complete elliptic integral of the first kind and $\pi$. By means of Mathematica we obtained numerical values for these calculated resistance as presented in Fig. 1 below.

The calculated resistance is plotted against the lattice site (2n,0,0), and from the figure shown it is cleared that the resistance in an infinite FCC lattice is approaching the finite value $f_o(3,0,0,0) = 0.4482203944$ as $n \to \infty$.

Figure 1 shows the resistance in an infinite FCC lattice against the site (*2n,0,0*) along the [100] direction. From this figure it is clear that the resistance is symmetric, and approaching a finite value.

A similar result was obtained for the resistance in an infinite SC network [35] where as the separation between the origin and any other lattice site the equivalent resistance approaches a finite value (i.e., $g_o = 0.505462$) which is the LGF at the origin in an infinite SC lattice, while the resistance in an infinite square network diverges for large separation between the two sites [36].

Table Caption

Table 1: Calculated values of F(2n,0,0;3)

Table 1:

| Site (2n,0,0) | R(2n,0,0) |
|---|---|
| (0,0,0) | 0 |
| (2,,0,0) | 0.371575 |
| (4,0,0) | 0.408775 |
| (6,0,0) | 0.421792 |
| (8,0,0) | 0.428366 |
| (10,0,0) | 0.432325 |
| (12,0,0) | 0.43497 |
| (14,0,0) | 0.43685 |
| (16,0,0) | 0.438396 |

Figure Caption

Fig. 1: Resistance R(2n,0,0) in an infinite FCC along the direction [1,0,0].

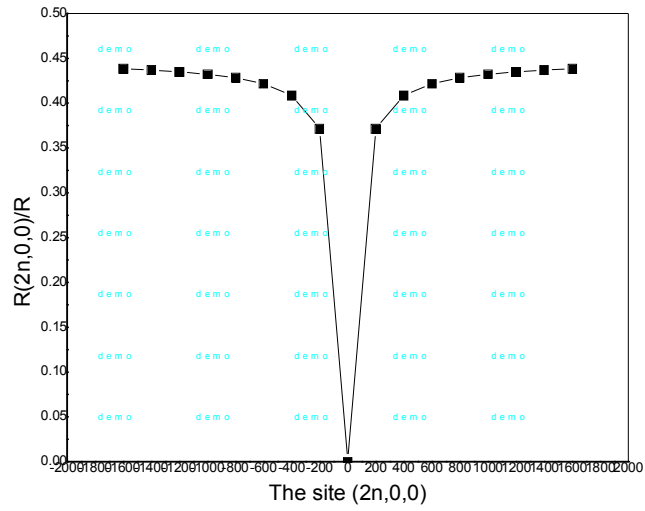

Fig. 1: